\documentclass{PoS}

\usepackage{epsfig,amsmath,times,color,cite,adjustbox,wrapfig,lipsum,booktabs,comment,textcomp}
\usepackage{url}
\usepackage[font=small,labelfont=bf]{caption}

\let\OLDthebibliography\thebibliography
\renewcommand\thebibliography[1]{
  \OLDthebibliography{#1}
  \setlength{\parskip}{0pt}
  \setlength{\itemsep}{0pt}
}

\newcommand\opt{-10pt}
\title{Prototype 9.7\,m Schwarzschild-Couder telescope for the Cherenkov Telescope Array: status of the optical system}

\ShortTitle{pSCT for CTA: status of the optical system}

\author{\speaker{D. Nieto}$^{1}$\footnote{Now at Departamento de F\'
    {i}sica At\'{o}mica, Molecular y Nuclear, Universidad Complutense
    de Madrid, 28040 Madrid, Spain}, T. B. Humensky$^{1}$,
  P. Kaaret$^{2}$, D. Kieda$^{3}$, M. Limon$^{1}$, A. Petrashyk$^{1}$,
  D. Ribeiro$^{1}$, J. Rousselle$^{4}$\footnote{Now at Subaru
    Telescope, John A. Burns Way, Hilo, HI 96720, USA},
  B. Stevenson$^{4}$, V. Vassiliev$^{4}$, P. Wilcox$^{2}$, for the CTA
  SCT Project\\

{\footnotesize
\\$^{1}$ Columbia University, Department of Physics,
$^{2}$ University of Iowa, Department of Physics and Astronomy,
$^{3}$ University of Utah, Department of Physics and Astronomy,
$^{4}$ University of California Los Angeles, Division of Astronomy and Astrophysics.}

E-mail: \email{nieto@nevis.columbia.edu}}

\abstract{The Cherenkov Telescope Array (CTA) is an international
  project for a next-generation ground-based gamma ray observatory,
  aiming to improve on the sensitivity of current-generation
  experiments by an order of magnitude and provide energy coverage
  from 30 GeV to more than 300 TeV. The 9.7m Schwarzschild-Couder (SC)
  candidate medium-size telescope for CTA exploits a novel aplanatic
  two-mirror optical design that provides a large field of view of 8
  degrees and substantially improves the off-axis performance giving
  better angular resolution across all of the field of view with
  respect to single-mirror telescopes. The realization of the SC
  optical design implies the challenging production of large
  aspherical mirrors accompanied by a submillimeter-precision custom
  alignment system. In this contribution we report on the status of
  the implementation of the optical system on a prototype 9.7\,m SC
  telescope located at the Fred Lawrence Whipple Observatory in
  southern Arizona.}

\FullConference{35th International Cosmic Ray Conference - ICRC2017\\
		10-20 July, 2017\\
		Bexco, Busan, Korea}

\begin{document}

\vspace{\opt}
\section{Introduction}
\vspace{\opt}
\label{sec:introduction}

The dual-mirror Schwarzschild-Couder (SC) optical design has been
shown to be a very promising solution for ground-based gamma ray
telescopes exploiting the imaging atmospheric Cherenkov technique
(IACT), providing a better angular resolution over a wider field of
view (FoV) than usual single-mirror
IACTs~\cite{2007APh....28...10V}. Three models of SC telescopes have
been proposed as candidates to form part of the next-generation IACT
observatory, the Cherenkov Telescope Array
(CTA\footnote{www.cta-observatory.org},~\cite{Acharya20133}). Two of
these models have been designed as small-sized telescopes (the 4.3\,m
aperture ASTRI telescope and 4.0\,m aperture GCT~\cite{CTA_SSTs}),
aiming to cover the high end of CTA energy band (>5 TeV). The
remaining model belongs to the medium-size class of CTA telescopes
(MST) and, with a 9.7 m diameter aperture, 8\textdegree\ field of view
(FoV), and 0.067\textdegree\ pixel size, it has been designed to fully
exploit the SC optical design in order to provide the best possible
angular resolution in the CTA core energy range (0.1 -- 10 TeV). This
contribution will describe the status of development of the optical
system for a prototype SC-MST (pSCT), currently under construction at
the Fred Lawrence Whipple Observatory (FLWO) in southern Arizona
(USA)\footnote{http://cta-psct.physics.ucla.edu/}. These proceedings
follow up on previous reports~\cite{2015ICRC-OPT,2015ICRC-AL}. A more
general description of the current status of the pSCT project as a
whole can be found elsewhere in these
proceedings~\cite{2017ICRC-pSCT}.

The structure of this work is as follows: a brief overview of the
optical system is provided in Section~\ref{sec:overview}; in
Section~\ref{sec:panels} we discuss the status of the mirrors;
Section~\ref{sec:alignment} contains a description of the alignment
system and a summary of the laboratory integration tests of the
different components this system is made of; a description of the
stray light and sunlight control system can be found in
Section~\ref{sec:stray}, followed by a summary and our outlook in
Section~\ref{sec:summary}.

\vspace{\opt}
\section{Overview of the optical system}
\vspace{\opt}
\label{sec:overview}

The advantages of the SC-MST optics design, as compared to
conventional Davis-Cotton or parabolic optics featured in all
current-generation IACTs, are numerous: it completely corrects
spherical and comatic aberrations, provides a fairly constant point
spread function over a large FoV, and allows a reduced plate scale,
permitting the implementation of finely pixelated focal plane
instrumentation. However, those benefits come at a cost; highly curved
aspheric mirror panels and tight alignment tolerances make the
implementation of a 9.7\,m aperture SC telescope a technological
challenge.

To minimize production costs, the 9.7\,m diameter primary mirror (M1)
is segmented into 48 mirror panels, split between two rings: an inner
ring (P1) of 16 panels and an outer ring (P2) of 32 panels. Similarly,
the 5.4-m-diameter secondary mirror (M2) is segmented into 24 mirror
panels, split between two rings: an inner ring (S1) of 8 panels and an
outer ring (S2) of 16 panels. Ray-tracing simulations of the SC-MST
optical system~\cite{2007APh....28...10V} show that, in order to
achieve a point spread function (PSF) compatible with the pixel size
of a high-resolution gamma-ray camera~\cite{2015ICRC-CAM} the
alignment precision must be at the sub-mm and sub-mrad levels, both
locally (panel-to-panel) and globally. In addition, the accuracy of
source localization requires 5 arcsec mirror tilt control. While these
tolerances remain very loose compared to the diffraction limit of a
similar-size optical telescope, they are far more strict than those of
current IACT optical systems and require automated mechanical
alignment and continuous monitoring of the segmented mirror surfaces.

The pSCT's alignment system consists of two major subsystems: a
\emph{global} alignment system (GAS) that continuously monitors the
relative positions of the main optical elements of the telescope (M1,
M2, and the camera focal surface). The GAS also monitors for
large-scale spatial perturbations of the M1 and M2 figures. The
\emph{panel-to-panel} alignment system (P2PAS) measures and corrects
for misalignments between neighboring panels and provides a continuous
monitoring of the alignment of the optical surfaces'
figures. Together, these systems will ensure the aforementioned sub-mm
and sub-mrad precision in the alignment and positioning of the pSCT's
main optical elements.

\vspace{\opt}
\section{Mirror panels}
\vspace{\opt}
\label{sec:panels}

\subsection{Primary mirror panels}
\vspace{-2mm}
\label{subsec:ppanels}

The primary mirror panels have characteristic sag of a few millimeters
and are produced using a cold glass slumping method. This process was
developed by Osservatorio Astronomico di Brera and Media Lario
Technologies (MLT), Italy. During the slumping process, two
1.7-mm-thick sheets of glass are assembled on each side of a 30-mm
aluminum honeycomb core and slumped to the required figure on a
precisely machined mandrel. M1 mirror panels were fabricated at MLT
and are then coated by Bte Bedampfungstechnik (BTE) in Germany. All P1
(19) and P2 (38) panels, including spares, for the primary mirror have
been fabricated, coated, and delivered to UCLA. Metrology data of the
P1 panels find an average slope error of 84$~\mu \textrm{rad}$ and an
RMS of 11.4$~\mu \textrm{rad}$, below the specification of 170$~\mu
\textrm{rad}$ and goal of 100$~\mu \textrm{rad}$. All but one panel
met the goal requirements.

Metrology data of the P2 panels find an average slope error of
151.2$~\mu \textrm{rad}$ and an RMS of 53.5$~\mu \textrm{rad}$. Seven
of the P2 panels meet the goal requirements, with 26 of them meeting
specification requirements.  Overall, the production process has
demonstrated success in manufacturing primary mirror panels within the
goal requirements; however, improvements are being made towards
industrialization of the fabrication process as the variables
affecting production quality are becoming better understood.

\begin{figure}[t]
  \centering
  \includegraphics[width=0.45\textwidth,clip=true,trim= 55 55 55 55]{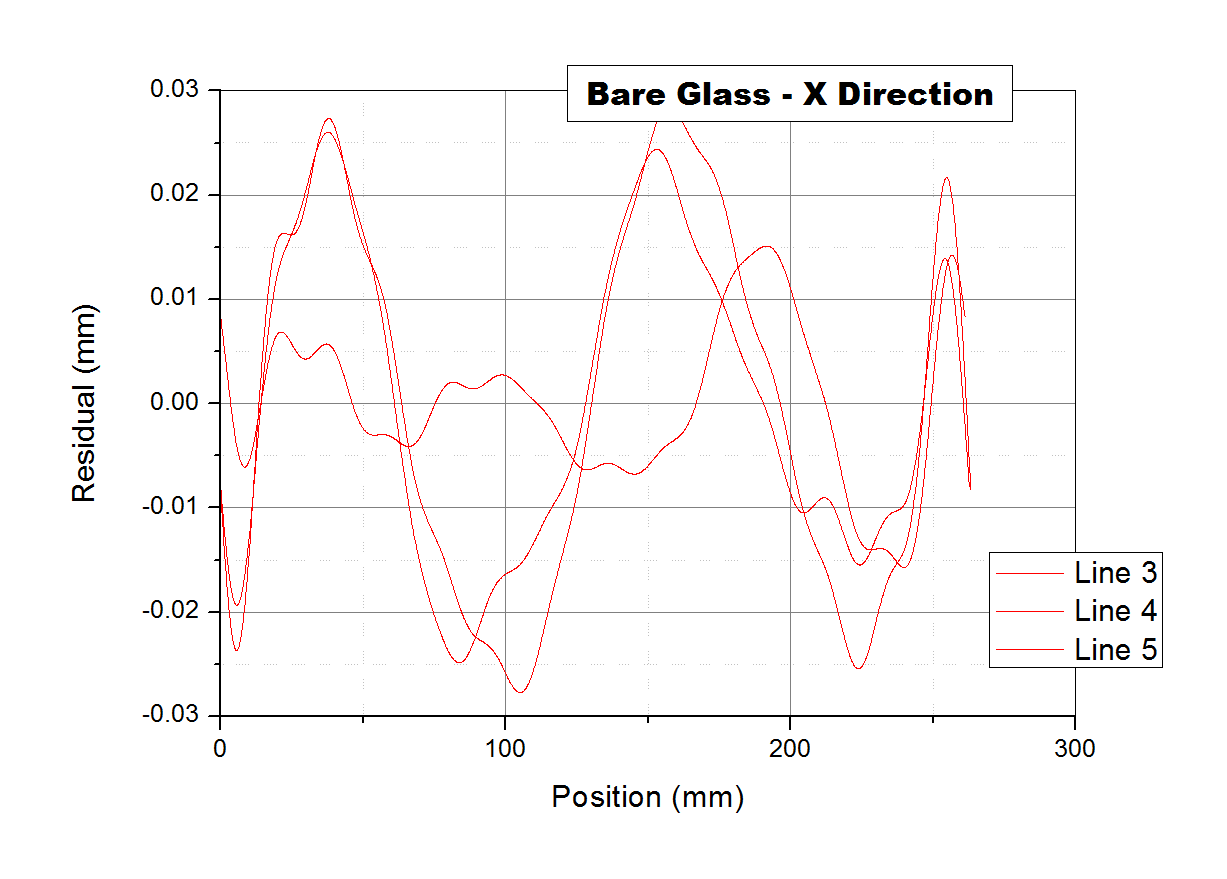}
  \includegraphics[width=0.45\textwidth,clip=true,trim= 55 55 55 55]{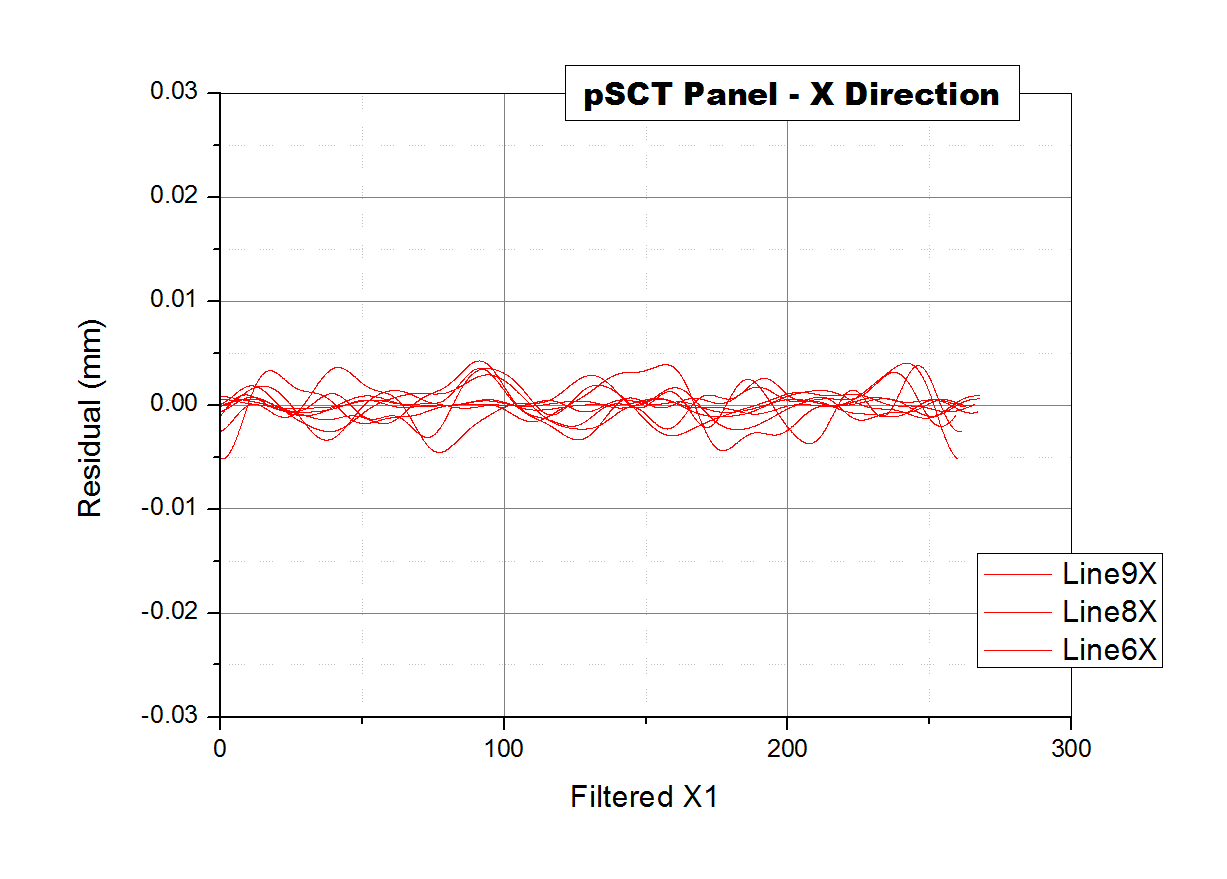}
  \caption{Example of residuals from the ideal figure of the optical
    surface of a S2 mirror panel. \emph{Left:} Residuals of the
    pre-shaped glass foil before assembly onto the final mirror panel
    (first production step). \emph{Right:} Residuals after assembly
    (second production step). All residuals are in millimeters.}
  \label{fig:residuals}
\end{figure}

\vspace{-3mm}
\subsection{Secondary mirror panels}
\vspace{-2mm}
\label{subsec:spanels}

The secondary mirror panels have characteristic sag of a few
centimeters and cannot be manufactured by cold slumping technology
alone. A two-stage process has been developed to fabricate the
secondary mirror panels, mitigating the main technical risk of the
pSCT project. First, thin (2 mm) glass foils are pre-shaped to the
figure of the highly curved S1 and S2 segments via hot glass slumping
replication on a precise low thermal expansion coefficient bending
tool. Flabeg, Germany, the developer of this technology, realizes this
fabrication step producing glass foils with residual deviations from
the ideal figure of$~\sim 0.3$-mm amplitude over a mid-scale spatial
period of 10-15 cm. Then, the pre-shaped glass foils are assembled
into hybrid S1 and S2 panels using the cold slumping method as in M1
panel production. This technology is know to correct large-scale
residual deviations; however, it can also achieve mid-scale fine
tuning of the panel figure, dramatically reducing such deviations, as
can be seen in Figure~\ref{fig:residuals}. Both S1 and S2 fine tuning
figuring mandrels have been produced at MLT and satisfy
specifications.  The production of M2 glass foils is completed at
Flabeg and their assembly into M2 panels is ongoing at MLT.  All M2
panels will be coated by Zaot, Italy.

The metrology data has been encouraging; around half of the S1 panels
satisfy specifications of $<200~\mu \textrm{rad}$, with the other half
just above this target. While the secondary mirror panel production
process is still being optimized and improvements are being made, it
is nonetheless a success that secondary mirror panels which satisfy
the SCT requirements have been produced at an acceptable cost for the
implementation of such a telescope design in the context of CTA.

\vspace{\opt}
\section{Alignment system}
\vspace{\opt}
\label{sec:alignment}

\subsection{Mirror panel modules}
\vspace{-2mm}
\label{subsec:p2pas}

\begin{figure}[t]
  	\centering
  	\includegraphics[height=4.6cm,clip=true,trim= 300 0 200 0]{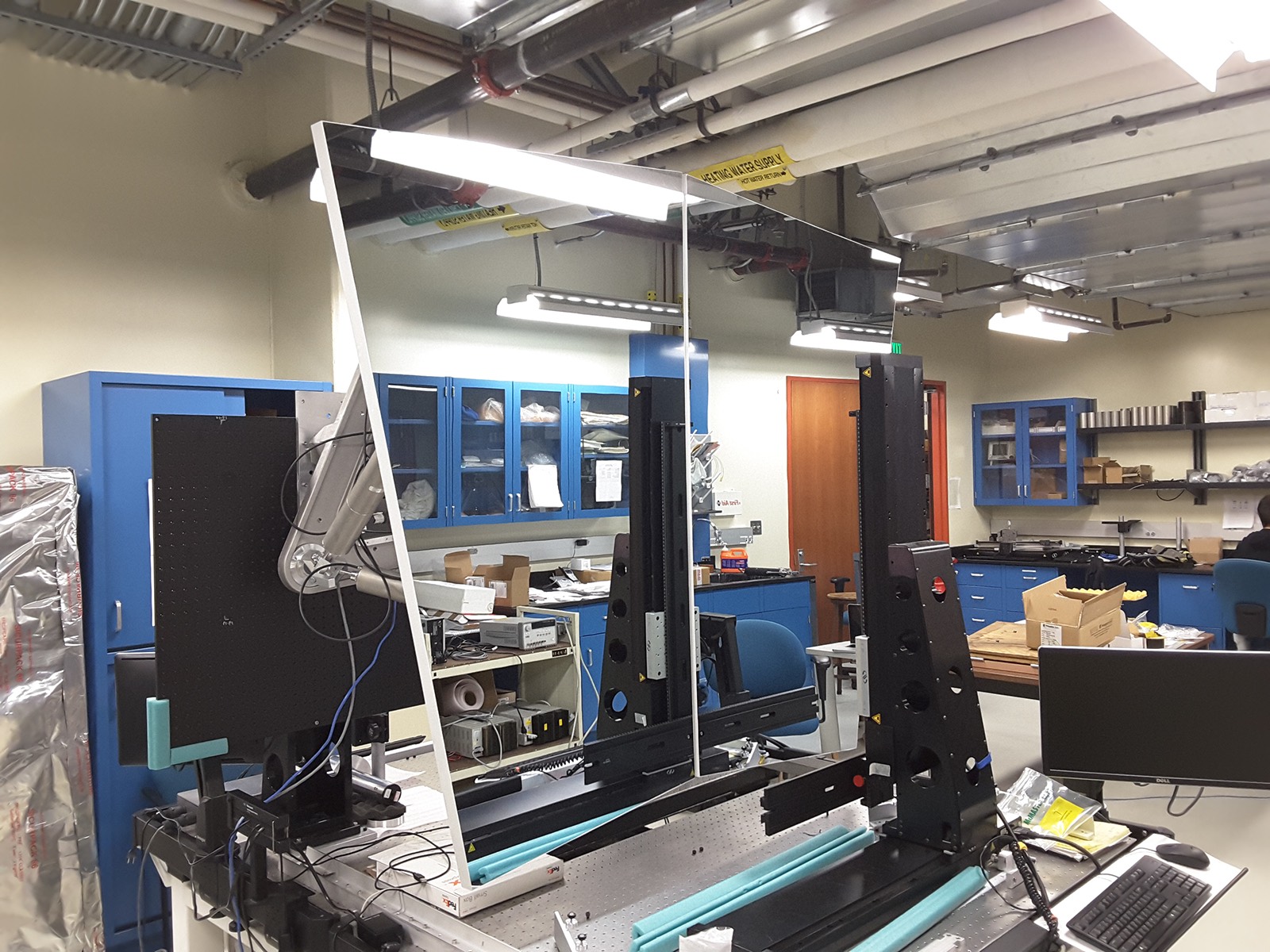}
  	\includegraphics[height=4.6cm,clip=true,trim= 0 0 0 0]{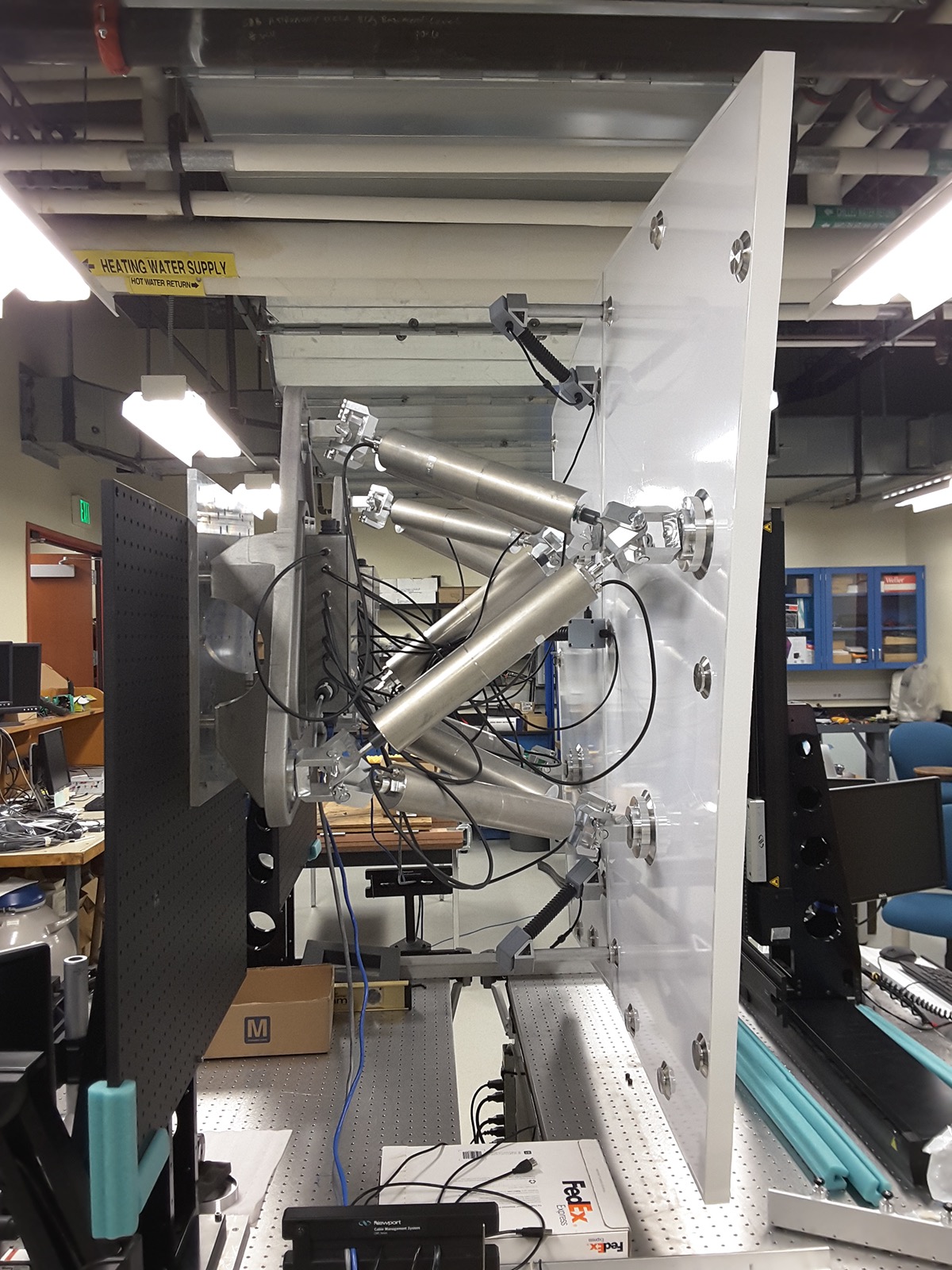}
  	\includegraphics[height=4.6cm,clip=true,trim= 0 0 0 0]{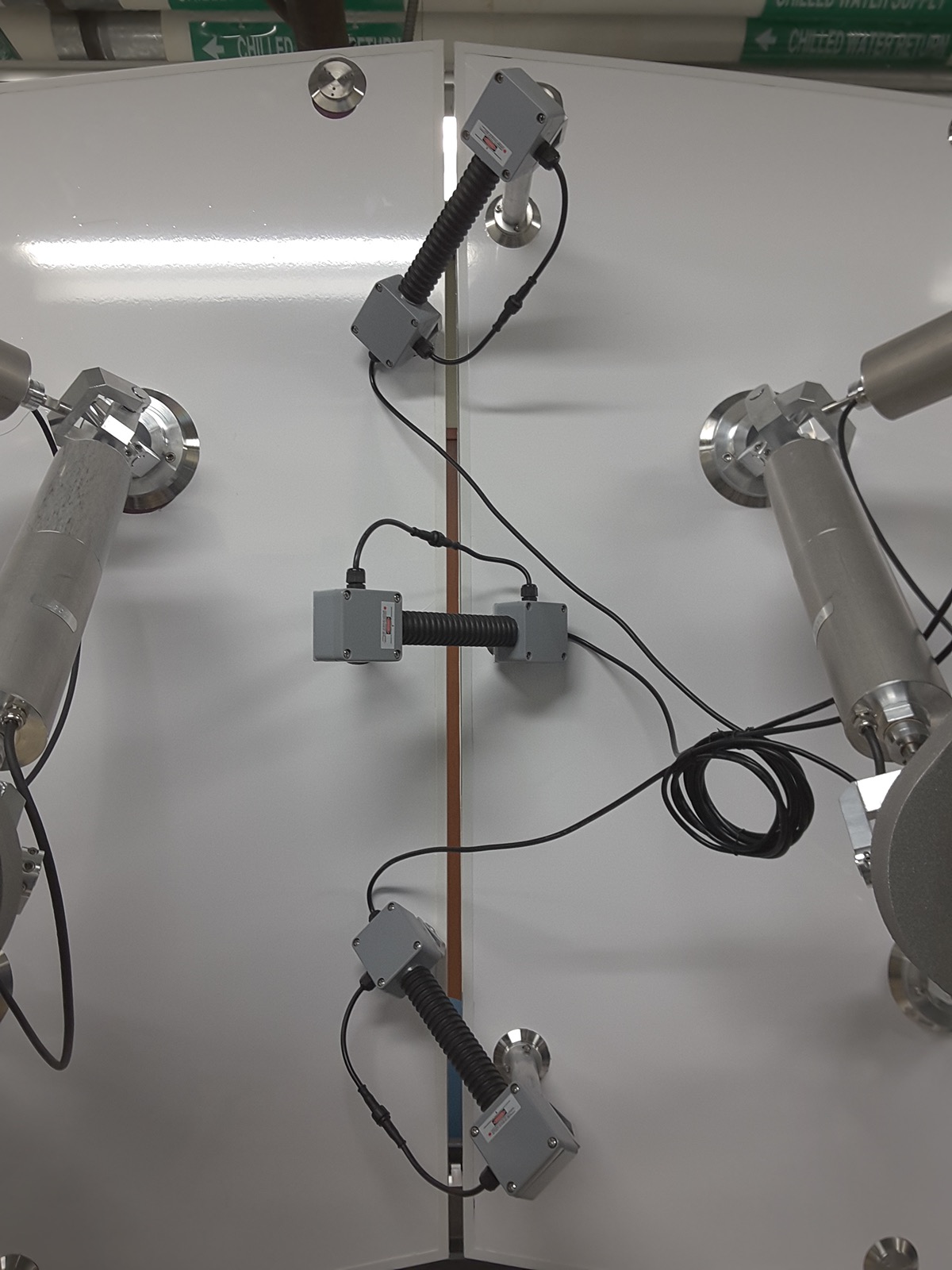}
  	\includegraphics[height=4.6cm,clip=true,trim= 0 0 0 0]{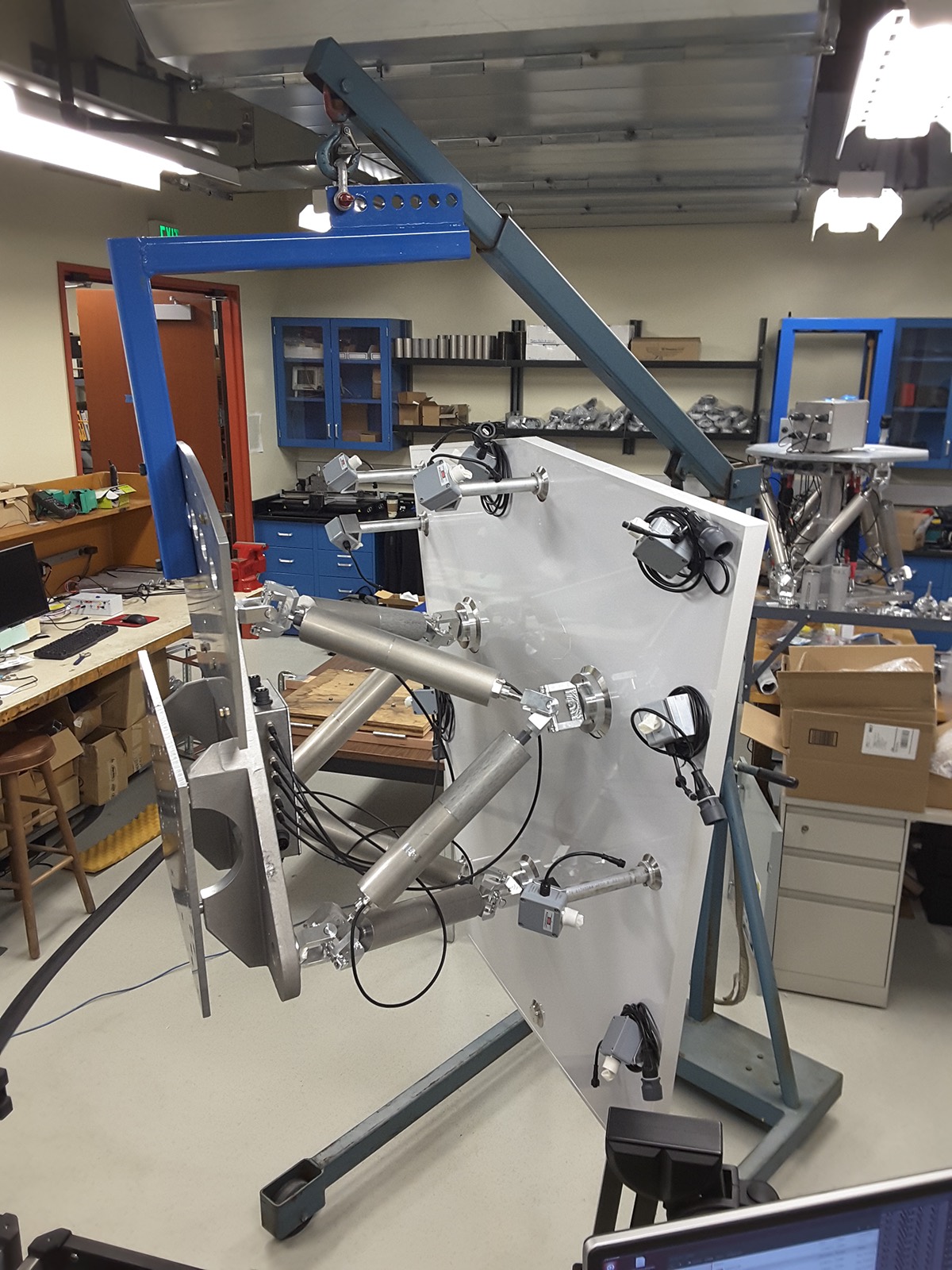}
	\caption{Photographs taken at the UCLA VHE Laboratory during
          the MPM integration tests. \emph{Far left:} Front view of
          two P1 MPMs installed onto the calibration set up, facing a
          coordinate measuring machine. \emph{Center left:} Lateral
          view of the same pair of P1 MPMs, showing the Stewart
          platform and attachment triangle. \emph{Center right:}
          Orthogonal triad of MPESs interfacing the two P1
          MPMs. \emph{Far right:} Complete P1 MPM attached to a
          lifting fixture.}
    \label{fig:mpm}
\end{figure}

The mirror panels, main component of the P2PAS, are attached to six
linear actuators arranged in a Stewart platform (SP) design, with the
relative positions between adjacent mirror panels being determined by
a collection of mirror panel edge sensors (MPESs). The conjunction of
SP and MPES measurements allows for panel-to-panel alignment with an
accuracy better than the required 100 $\mu m$. Control of the SP and
MPES for each mirror panel, as well as measuring of the external
temperature of each SP assembly, is provided by an associated mirror
panel controller board (MPCB), mounted to an aluminum triangle which
interfaces the mirror panel with the optical support structure
(OSS). A mirror panel module (MPM) is defined to be a mirror panel
with its accompanying SP, MPES, MPCB, and mounting triangle.  A
central computer in charge of pSCT alignment controls all 72 MPMs
through the MPCBs with a minimal refresh period of measurement and
position adjustment of$~\sim 15$ seconds per cycle.  A more detailed
description of the MPM elements is given in the previous ICRC
proceedings~\cite{2015ICRC-OPT}.

\begin{figure}[t]
  \centering
  \includegraphics[width=0.42\textwidth,clip=true,trim= 15 15 15 15]{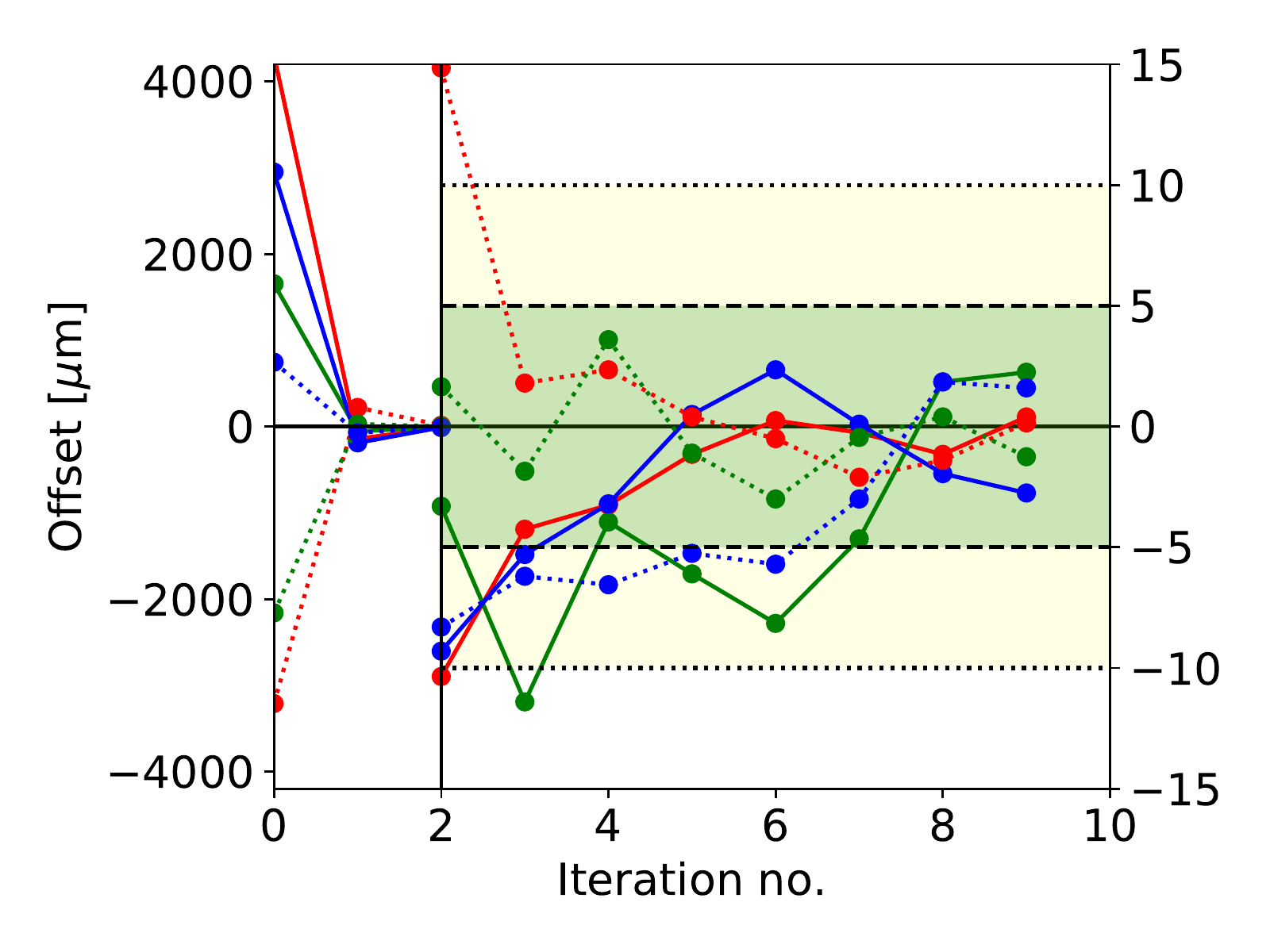}
  \quad
  \includegraphics[width=0.42\textwidth,clip=true,trim= 15 15 15 15]{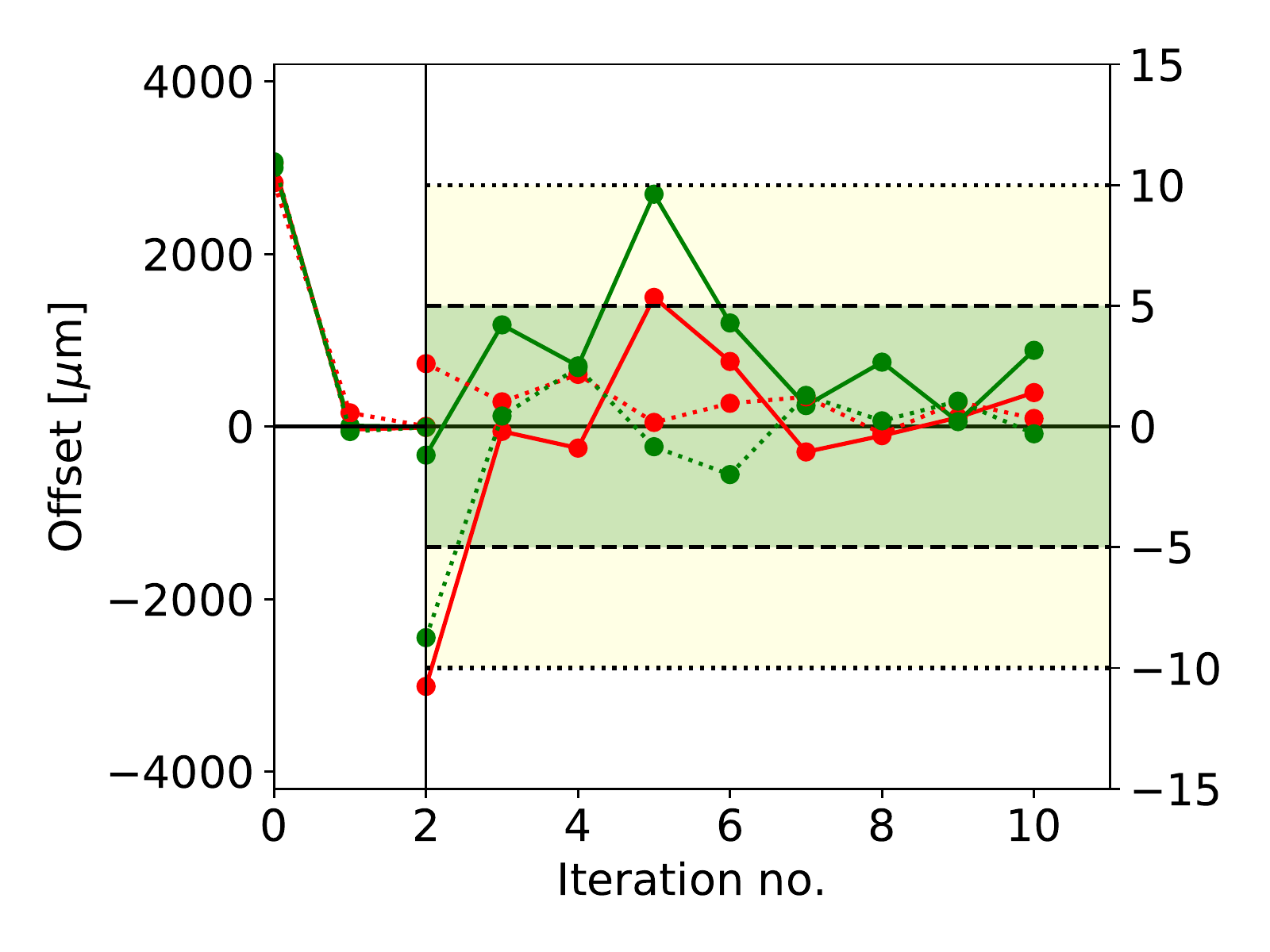}
  \caption{Examples of alignment convergence tests. The vertical line
    marks the change of scale for the Y axis.  \emph{Left:} Readings
    from a triad of MPES interfacing two P1 panels (as in
    Fig.~\ref{fig:mpm}, center right) during a MPM alignment
    convergence test. Solid (dotted) lines represent the offset from
    the target position in the X (Y) MPES local coordinate system.
    \emph{Right:} Readings from the two PSDs located in the secondary
    mirror OT.}
  \label{fig:conv}
\end{figure}

All elements of the MPMs have been assembled, tested, and
calibrated. This includes 480 actuators (and accompanying 4
d.o.f. joints, which interface the actuators to the mirror panels and
triangles), 80 MPCBs and 380 MPES.  The calibration setup for the MPMs
has been successfully established at the UCLA Very High Energy (VHE)
Laboratory (see Figure~\ref{fig:mpm}), where an alignment of under
$5~\mu m$ between mirror panels, which translates into an alignment
under $10 \mu \textrm{rad}$, has been demonstrated as shown in
Figure~\ref{fig:conv} (left panel).

\vspace{-4mm}
\subsection{Global alignment}
\vspace{-2mm}

The global alignment system measures the relative positions of the
pSCT's primary and secondary mirrors and camera.  To make the
contribution of global alignment errors negligible in the overall
error budget and to meet post calibration pointing specifications, the
position of pSCT optical system elements needs to be known with the
following accuracy: for the camera center $\pm 0.27$ mm ($\pm 10$
arcsec on sky) perpendicular to the optical axis ($\Delta x$, $\Delta
y$), for the camera focal surface $\pm 1.2\ \textrm{mm}$ along the
optical axis ($\Delta z$) and for the M1 position $\pm
7.0\ \textrm{mm}$ along the optical axis ($\Delta z$). The angular
requirements are $\pm 0.05\ \textrm{mrad}$ ($\Delta \theta_x$, $\Delta
\theta_y$) ($\pm 10$ arcsec on sky). The output of the global
alignment system can be fed back into the mirror active control (see
Section \ref{subsec:p2pas}) to correct for deformations due to
temperature and gravity, with a minimal alignment refresh period of
tens of seconds. The alignment system design was driven by the
requirement that adjustments will be needed due to significant
temperature variations, which cause deformations in the OSS from
season-to-season or possibly as frequently as night-to-night. The
rigid design of the pSCT mechanical structure is expected to produce
only small deformations caused by varying gravitational load which
which should not trigger OS re-alignment. The current implementation
of the alignment system design can potentially allow for re-alignment
of the pSCT optics every run (typically 20-30 minutes). The needed
refresh rate of the alignment system will be studied on the pSCT for
future corrections in SCT implementation for CTA. We anticipate
running the global alignment system continuously to monitor the
positions of selected mirror panels every 2 seconds.

\begin{figure}[t]
  \centering
  \includegraphics[width=0.49\textwidth,clip=true,trim= 0 0 0 0]{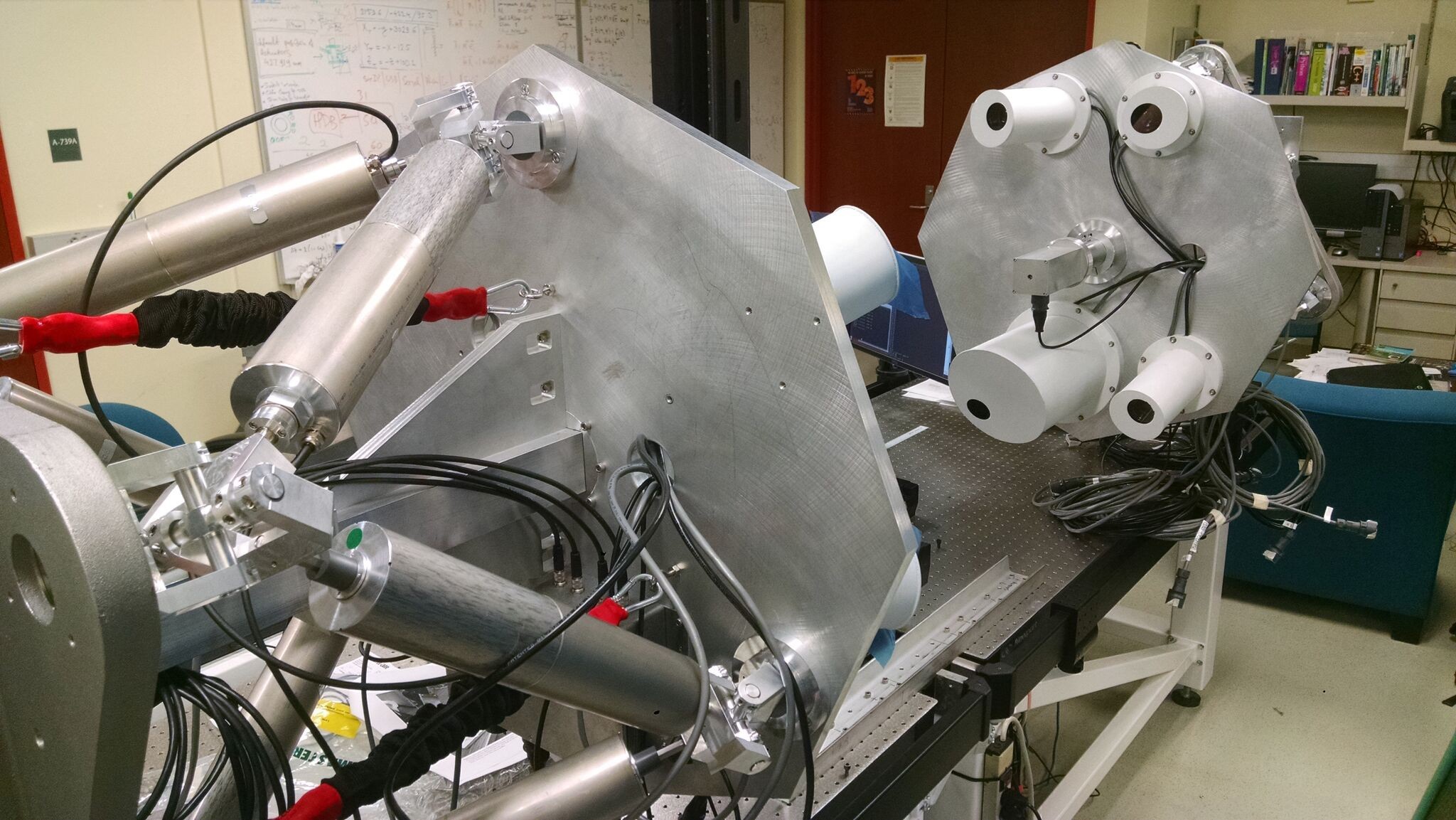}
  \includegraphics[width=0.49\textwidth,clip=true,trim= 0 0 0 0]{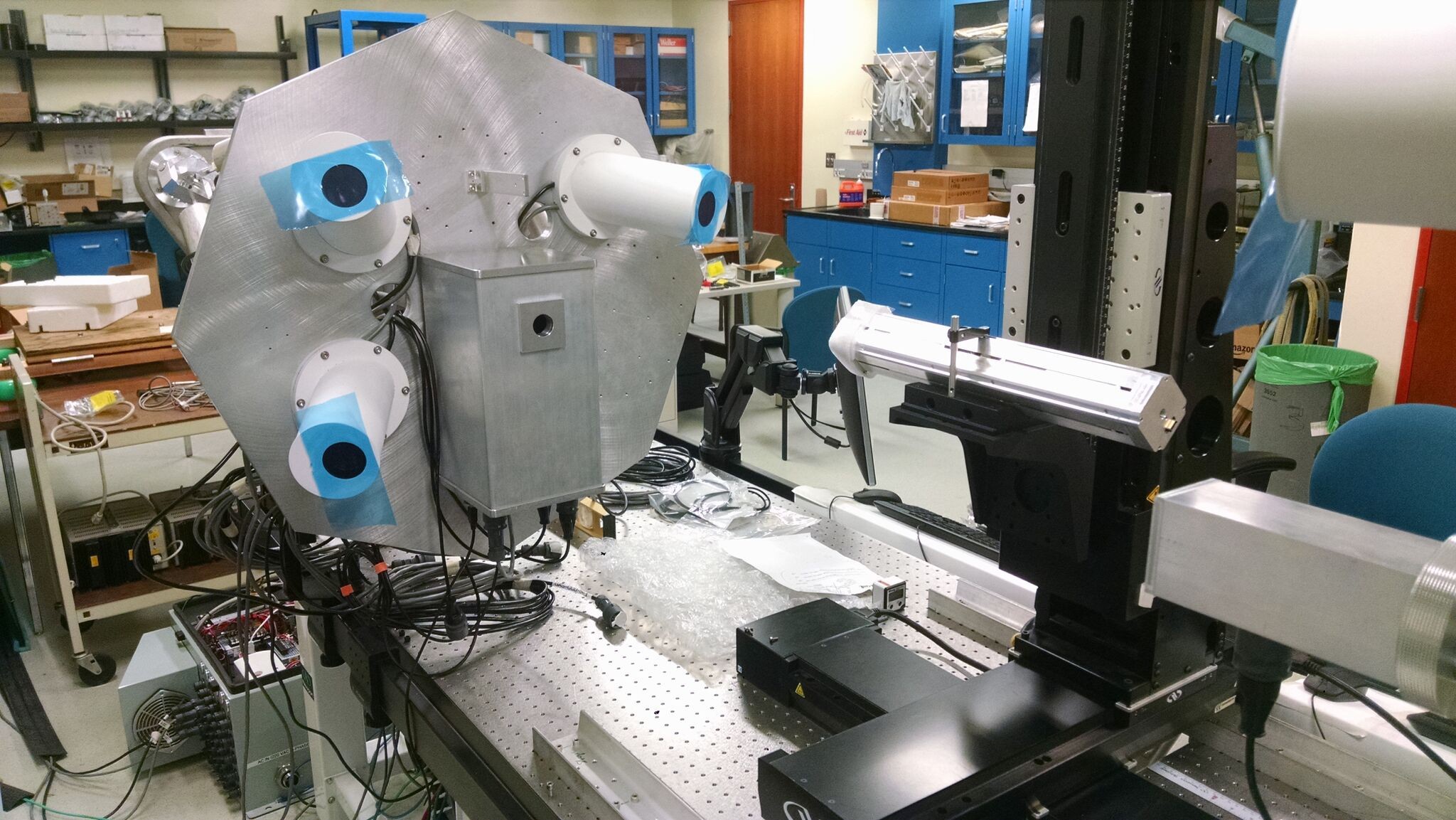}
  \caption{Photographs taken at the UCLA VHE Laboratory during
    the OTs integration tests. \emph{Left:} Rear view of the
    secondary OT, facing the primary OT. \emph{Right:} Front
    view of the secondary OT facing the PSD-equipped camera
    module for initial focal plane alignment with the optical
    axis of the telescope.}
  \label{fig:ots}
\end{figure}

During telescope operation, the global alignment is measured by two
different systems: LED / CCD pairs to measure translation and
auto-collimator/reflector pairs to measure angular change.  Six LEDs
are mounted in rectangular patterns on three different panels per
mirror and eight LEDs are mounted on an octagon attached to the focal
plane of the gamma-ray camera. The CCDs are mounted on two optical
tables (OTs) positioned in the centers of M1 and M2 along the
telescope's optical axis, with each CCD having a fixed pointing
towards its paired LED targets (7 LED/CCD pairs total). There is one
auto-collimator/reflector pair per mirror, also installed on the OTs,
and it is pointed at one of the panels also being monitored by a
LED/CCD pair.  The OTs are central to the global alignment of the
telescope. Each OT is made up of a Stewart platform, with an aluminum
plate mounted on one end and the other end mounted to the OSS, located
on the telescope's optical axis. The M1 OT houses a laser, and the M2
OT houses a rangefinder, a pair of position sensitive devices (PSDs),
and a sky camera. First, the M1 OT is adjusted such that the laser
points along the mechanical axis of the telescope. Then the M2 OT is
adjusted, using the PSD and rangefinder, such that it moves into the
calibrated aligned position with respect to the M1 OT. Once the OTs
are aligned, the CCD/LED process can begin to achieve global alignment
between M1 and M2. A module containing two PSDs has also been built
and calibrated, which can temporarily replace the central camera
module and be used to calibrate the camera plane with respect to the
optical axis (see Figure~\ref{fig:ots}, right panel).

The OTs are completely assembled, and the calibration procedure for
the OTs has been established at UCLA (see Figure~\ref{fig:ots}). This
calibration procedure demonstrated the alignment of the laser position
on the PSDs to have an accuracy of better than 10 $\mu$m (for a
calibration distance of $\sim$ 2 m), corresponding to an angle below
the mrad level on the telescope (see Figure~\ref{fig:conv} (right
panel). During lab testing of prototype LED/CCD equipment using the
expected CCD to gamma ray focal plane distance of 1.9 m, we measured a
perpendicular accuracy of $\Delta x$, $\Delta y\ <\ 0.03\ \textrm{mm}$
and on-axis accuracy of $\Delta z\ <\ 0.06\ \textrm{mm}$. Using a
nominal CCD-to-panel distance of 8.0 m we measured a perpendicular
accuracy of $\Delta x$, $\Delta y\ <\ 0.04\ \textrm{mm}$ and on-axis
accuracy of $\Delta z\ <\ 1.14\ \textrm{mm}$. The angular accuracy of
the auto-collimator/reflector system has a measured angular accuracy
of $\Delta \theta_x$, $\Delta \theta_y\ <\ 0.016\ \textrm{mrad}$. The
lab tests are all well within tolerances.

\vspace{-3mm}
\subsection{Power and connectivity}
\vspace{-2mm}
\label{subsec:pedb}

The power and connectivity served to all MPMs and OTs are provided by
a pair of power and Ethernet distribution boxes (PEDBs), one per
mirror, which will be installed onto the telescope M1 and M2 OSS. The
PEDBs are equipped with a number of 24 VDC power supplies and 1G
Ethernet switches controlled by a low-power ARM computer also
operating the environment control (air cooling). Voltages from all
power supplies and currents in all SPs are monitored. Both PEDBs have
been fully assembled and all ports have been calibrated for voltage
and current readings, as well as the internal temperature sensors, and
are fully operational.

\vspace{-3mm}
\subsection{Alignment control software}
\vspace{-2mm}
\label{subsec:acs}

The alignment software performs monitoring of misalignment between
MPMs and attempts mirror alignment when necessary. As described
in~\cite{2015ICRC-AL}, the software is implemented through the OPC-UA
communication protocol, following a server-client paradigm, with each
MPCB acting as a server of MPES data and actuator control. To allow
for failing components and their easy replacement, and extensions of
the alignment system in the future, the software addresses hardware
components through generic interfaces, retrieving initialization
information from a database. This use of generic interfaces also lets
us treat global alignment components in the same way as MPMs, feeding
coordinate data from CCDs and PSDs into the alignment procedure. The
central computer exposes all the necessary alignment data and methods
through a top-level OPC-UA server, enabling straightforward
integration into CTA's Array Control.

To be more precise, the alignment software solves a minimization
problem, attempting to keep P2PAS and GAS readings at their nominal
values. Introducing a panel's misalignment parameter $\chi^2_m =
\chi^2_{m,\text{P2PAS}} + w_{GAS}\chi^2_{m,\text{GAS}}$, where
$\chi^2_{m,\text{P2PAS}}$ is its misalignment coming from P2PAS
readings, $\chi^2_{m,\text{GAS}}$ comes from global misalignment, and
$w_{GAS}$ is a weight that effectively forces treating global
misalignment more (or less) stringently than panel-to-panel
misalignment, this can be written in simplified notation as minimizing
\[ \sum_m\chi^2 =\sum_m\left[ \sum_{\text{P2AS}}\left(\delta \vec s_m (\text{P2PAS}) + \hat R_m\delta \vec L_m\right)^2+ w_{\text{GAS}}\sum_{\text{GAS}}\left(\delta \vec s_m (\text{GAS}) + \hat M_m\delta \vec L_m\right)^2\right].\] 
Here, for a given panel $m$, $\delta \vec s_m$ are its positional
readings coming from edge sensors (P2PAS) or the global alignment
system (GAS); $\delta \vec L_m$ are the actuator displacements that
bring it to its nominal position -- these are to be solved for; $\hat
R_m$ and $\hat M_m$ are the actuator response matrices that connect
positional readings with physical motions of actuators. This form
highlights the robustness of the alignment system -- even with some
hardware failing, $\chi^2$ can still be minimized to a satisfactorily
small value.  The weight parameter has not been set yet, and will be
tuned empirically. The ideal response matrices $\hat M_m$ are
identical for all panels, and the real response matrices are only
small deviations from this ideal matrix.  The response matrices $\hat
R_m$, however, are different for each pair of mirror panels and need
to be measured in lab.

The alignment software code of the central computer is multithreaded,
collecting sensor data $\delta \vec s_m$ and requesting actuator
movements by necessary displacements $\delta \vec L_m$ simultaneously
for each individual MPM. This results in the whole mirror aligning
nearly as fast as an individual panel.

\vspace{-4.5mm}
\section{Sunlight and stray light control}
\vspace{-3mm}
\label{sec:stray}

A specific feature of the dual-mirror pSCT is the need of the sunlight
protection system during day time parking. Such a system has been
implemented as a set of baffles, one per mirror, to block most of the
sunlight entering the telescope and also being reflected from
it. These baffles, along with three different parking positions that
alternate throughout the year, provide a safe solution to the sunlight
concentration problem as extensively modeled by ray-tracing
simulations. In addition, the specific dimensions and arrangement of
the baffles have been chosen so to minimize the amount of stray light
reaching the gamma-ray camera during normal operations.

\begin{wrapfigure}{l}{0.4\textwidth}
\vspace{-0mm}
  \centering
  \includegraphics[width=0.4\textwidth,clip=true,trim= 100 100 700 300]{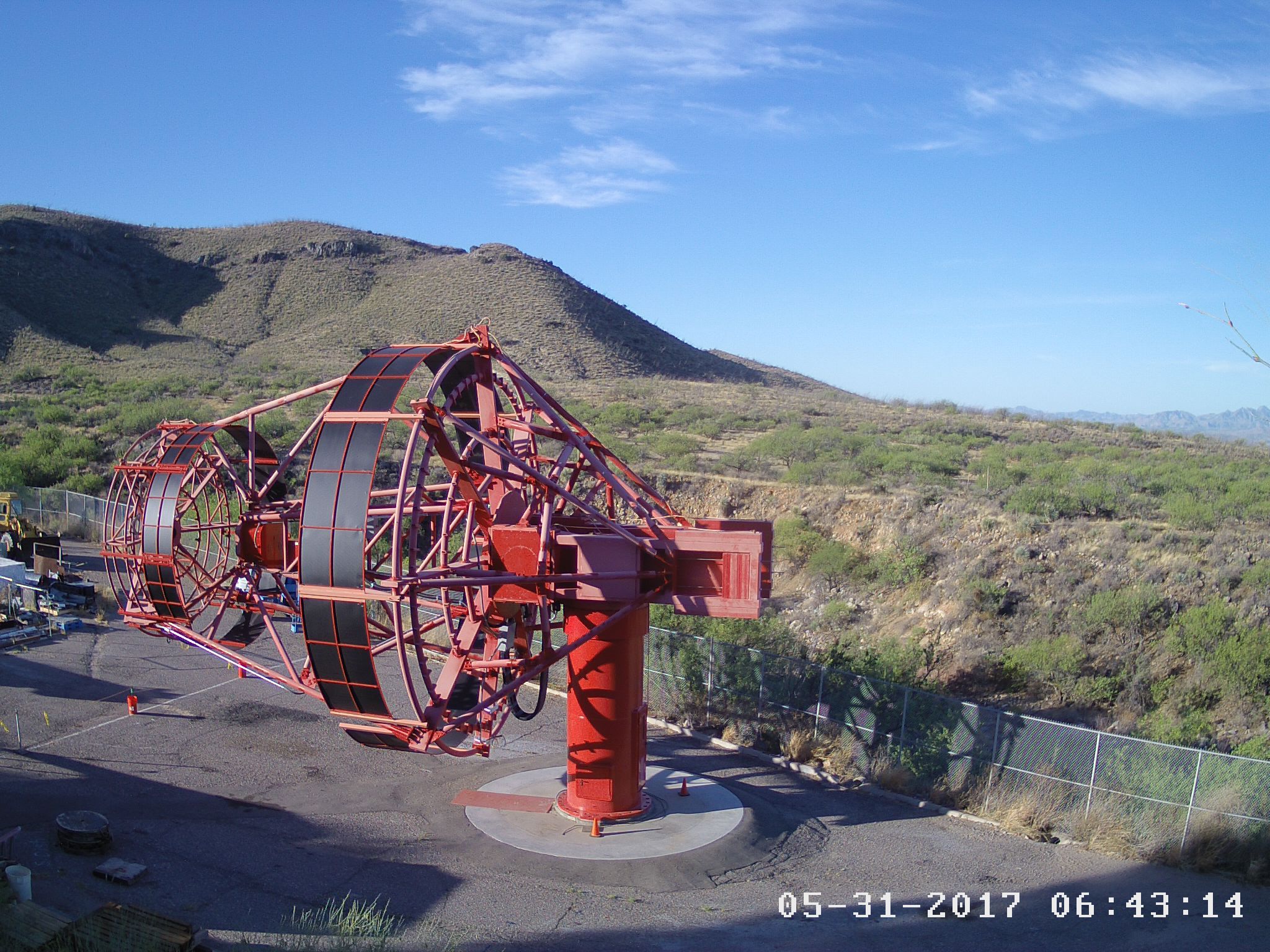}
  \caption{Prototype SCT, as of June 2017, showing both the primary and secondary baffles already installed.}
\vspace{-7mm}
\label{fig:webcam}
\end{wrapfigure}

The primary and secondary baffles were constructed in Spring 2016, and
were successfully fitted and installed onto the pSCT in Spring 2017
(see Fig.~\ref{fig:webcam}).

\vspace{\opt}
\vspace{-1mm}
\section{Summary and Outlook}
\vspace{\opt}

We have presented the current status of the optical system of the
prototype Schwarzschild-Couder telescope, currently under construction
at the Fred Lawrence Whipple Observatory. All mirror panels for the
telescope's primary mirror have been delivered to UCLA VHE Laboratory.
It has been proven that the highly-curved mirror panels for the
secondary mirror can be fabricated to meet their corresponding
requirements, and production process optimization is
underway. Integration tests for the MPMs and OTs have been
satisfactorily conducted, and alignment accuracies below $10\ \mu$m
have been consistently achieved. The power and connectivity of all
alignment components is guaranteed by the two calibrated PEDBs, and
their integrated operation by a mature alignment control software. The
sunlight and straylight control elements have been already installed
onto the telescope. Integration of the remaining components of the
optical system with the telescope structure is expected to begin in
July, 2017, with the installation of the OTs. Installation of the
mirrors, beginning with the primary, will follow and extend into the
early fall, allowing commissioning of the telescope to begin as the
summer monsoon season is ending.

\label{sec:summary}
\vspace{\opt}
\section{Acknowledgments}
\vspace{\opt}

This work was conducted in the context of the SCT project of the CTA
Consortium. The project has been made possible by funding provided
through the U.S. National Science Foundation Major Research
Instrumentation program (MRI award 1229792) and from the agencies and
organizations listed here: {\small
  \url{http://www.cta-observatory.org/consortium_acknowledgments}},
which we gratefully acknowledge.  We particularly recognize the
exceptional contributions to the project made by the technical support
staff at the Fred Lawrence Whipple Observatory.
\vspace{\opt}

\end{document}